# A Method for Constructing Minimally Unsatisfiable CNFs
Robert Cowen[1]
Mathematics Department
Queens College, CUNY

## 1. Introduction

Ivor Spence[1] has developed an ingenious method for easily generating unsatisfiable 3-cnfs that turn out to be rather difficult for ATPs (automated theorem provers). In this paper, we generalize his construction to cnfs of arbitrary clause length and then show that the unsatisfiable cnfs generated are, usually, "minimally unsatisfiable," that is, the removal of even one clause results in a satisfiable cnf. We first review Spence's method in a more general setting and then illustrate this minimality property.

## 2. Preliminary Definitions

A *literal* is a propositional variable or its negation. A *clause* is a disjunction of literals and a *conjunctive normal form* or *cnf* is a conjunction of clauses. A *k-cnf* is a cnf where all it clauses have exactly $k$ literals. A cnf is *satisfiable* if there is an assignment of truth values, t, f, to its propositional variables that makes it true (t) when evaluated using the usual truth-table rules.

## 3. Unsatisfiable CNF Construction

Given positive integers $g, k$, suppose the $(2k - 2)g + 1$ propositional variables, $p_1$, $p_2$, ..., $p_{(2k-2)g+1}$, are partitioned, in order, into $(g -1)$ sets of size $(2k - 2)$ and one set of size $(2k - 1)$. For each cell of the partition form all k-clauses from the variables in that cell and let $C_1$ be the conjunction of all these k-clauses. If $C_1$ is to be satisfiable no more than $(k - 1)$ variables from each partition cell can be false; thus no more than $(k - 1)g$ variables can be false.

Next let $q_1$, $q_2$, ..., $q_{(2k-2)g+1}$, be a random permutation of the $p$'s and again partition, in order, into $(g -1)$ sets of size $(2k - 2)$ and one set of size $(2k - 1)$. This time, for each cell of the partition, form all k-clauses from the negated variables in that cell and let $C_2$ be the conjunction of all these k-clauses. If $C_2$ is to be satisfiable, no more than $(k - 1)$ variables from each partition cell can be true; thus no more than $(k - 1)g$ variables can be true.

Let $C$ be $C_1 \wedge C_2$. If $C$ is to be satisfiable, both $C_1$ and $C_2$ must be satisfiable; thus no more than $(k - 1)g$ variables can be false and $(k - 1)g$ variables can be true. However, $(k - 1)g + (k - 1)g = (2k - 2)g < (2k - 2)g + 1$ ! Thus $C$ is an unsatisfiable cnf.

## 4. Minimally Unsatisfiable CNFs

Suppose next that we drop one of the clauses in $C$, say for example, $p_1 \vee p_2 \vee ... \vee p_k$; let $C_1^{'} = C_1$ minus the clause $(p_1 \vee p_2 \vee ... \vee p_k)$. Let $C_2^{'} = C_2$, and $C^{'} = C_1^{'} \wedge C_2^{'}$. Let $\sigma$ be an assignment that assigns the truth values false to $p_1$, $p_2$, ..., $p_k$ and true to the remaining variables in the first p-cell. As long as no more than $(k - 1)$ propositional variables in each of the remaining cells of the partition of the $p$'s are assigned the value false, $C_1^{'}$ would be true under $\sigma$. Whether or not $C_2^{'}$ and hence $C^{'}$ is true under $\sigma$ will depend on whether $\sigma$ also has the property that at most $(k - 1)$ proposi-

$$C_1^{'},$$

$$C^{'}$$

$$C^{'} \qquad\qquad p_1,\ p_2,\ ...,\ p_k,$$

tional variables in each cell of the partition of the randomly permuted variables (the $q$'s) are assigned the value true under $\sigma$. While this is unlikely for any given assignment, there are so many assignments $\sigma$ satisfying $C_1'$, it is very likely that **some** assignment will have this property and the reduced cnf, $C'$, will then be true. It is this intuitive argument we shall investigate below. Note that for $C'$ to be true, the $k$ variables, $p_1, p_2, ..., p_k$, can now be false in the first cell, as long as each of the remaining $(g$ -1) cells in the $p$-partition have at most $(k$ -1) false variables; thus $(k - 1) g + 1$ variables can be false and, as above, $(k - 1)g$ variables can be true. However, $(k - 1) g + 1 + (k - 1)g = (2k - 2)g +1$ and the argument showing $C$ to be unsatisfiable cannot be applied to $C'$.

### 5. Experimental Results

The table below summarizes computer experiments we have conducted to test our supposition that most of the formulas constructed above are minimally unsatisfiable (MU). Each line in the table represents results on 500 formulas. So, for example, the first line constructs 500 3-cnfs, based on $4(5) + 1 = 21$ variables; there are $8(5) + 12 = 52$ clauses in each.

It turned out that 65% were MU. We also kept track of the number of times deleting a single clause in a formula resulted in a satisfiable subformula; we refer to this number as the *satisfiable number* of the formula. Thus, for an MU formula, its satisfiability number equals the number of its clauses. Even formulas that weren't MU were "almost MU," as shown by the fact that the Mean of the satisfiablility numbers(Mean Sat. No.) of the tested formulas is close to the number of clauses. We also include the Standard Deviation of these satisfiability numbers.

| k | g | clause number | MU Percent | Mean Sat. No. | Standard Dev. |
|---|----|---------------|------------|---------------|---------------|
| 3 | 5  | 52            | 65         | 51.3          | 0.71          |
| 3 | 8  | 76            | 79         | 75.6          | 0.89          |
| 3 | 10 | 92            | 82         | 91.6          | 0.89          |
| 3 | 12 | 108           | 83         | 107.7         | 0.71          |
| 3 | 15 | 132           | 88         | 131.8         | 0.66          |
| 4 | 5  | 190           | 82         | 189.7         | 0.93          |
| 4 | 6  | 220           | 88         | 219.8         | 0.82          |

### 6. Conclusion

The method outlined above provides an easy way to obtain MU formulas since such a large percentage of the formulas obtained by Spence's method are MU and it is easy to determine whether or not a given unsatisfiable formula is MU. Also we conjecture that the MU percentages approach 100% for each k, as $g \rightarrow \infty$. More experiments can be carried out on faster computers (we only used a laptop and a desktop) to test this conjecture. Hopefully a theoretic argument can be given to settle this conjecture.

We wish to thank Professor Lenore Cowen, Tufts University and Professor Emeritus, Peggy Strait, Queens College, CUNY for their valuable advice in the course of this work.

[1]Professor Emeritus.